# EVALUATING A TRANSITION-JUMP SYSTEM FOR THE FERMILAB MAIN INJECTOR USING XSUITE

A. P. Schreckenberger*, R. Ainsworth, M. Xiao, Fermilab, Batavia, IL, USA

*Abstract*

We describe an Xsuite simulation framework for the Fermilab Main Injector (MI) along with an evaluation of transition-crossing behaviors in the accelerator. In particular, we studied the introduction of quadrupole magnets into the lattice as part of a transition-jump system that will be implemented through the $2^{nd}$ Proton Improvement Plan (PIP-II). Simulated beam losses spurred by transition-induced instabilities were assessed under several systematic effects, including MI quad errors, magnet-to-magnet variability in the jump magnets, power-supply errors, and timing jitter.

## OVERVIEW

Accelerating particles through transition has yielded well-documented challenges that necessitated both technological developments and the consideration of instabilities [1–3]. Commonly referenced by its Lorentz factor $\gamma_t$, an accelerator's transition is set by the lattice. In the Fermilab Main Injector (MI), $\gamma_t \approx 21.6$ and remains static.

Beam behaviors rapidly change as the ramp nears transition, particularly with regard to the longitudinal dynamics, and circular accelerators must quickly shift configuration conditions in order to maintain particle storage. In the MI, flipping the RF phase from $\phi_s$ to $\pi - \phi_s$ and adjusting the chromaticity are the primary vehicles for retaining stability.

Despite these controls, measurable detriments manifest near the time of the transition crossing ($t_{\gamma_t}$), and simulations demonstrate both distortions of the longitudinal phase space and induced beam losses. Figure 1 (*left*) shows the longitudinal phase-space distribution from Xsuite [4] for the MI lattice at 5.25 ms before transition. Comparing this shape to Fig. 1 (*right*), which depicts the distribution near transition, highlights some of the impacts on beam behavior; the bunch length is spatially compressed, and $\delta = \Delta p/p$ is smeared.

In the MI, instabilities and beam losses driven by nonadiabatic motion and chromatic nonlinearities are the primary concerns [5]. Both of these instabilities emerge in time windows around transition, $t_{\gamma_t} \pm T_c$, where $T_c$ refers to either nonadiabatic ($T_{na}$) or nonlinear ($T_{nl}$) considerations. Within $t_{\gamma_t} \pm T_{na}$, particles are unable to keep up with the rapidly changing bucket shape. Meanwhile, chromatic nonlinearities induce the Johnsen effect, which generates a non-constant $\gamma_t$ for the particle ensemble. From the functional forms of $T_{na}$ and $T_{nl}$, we expect $T_{na} \approx 1.819$ ms and $T_{nl} \approx 1.113$ ms in the accelerator. Using Eqs. (1) and (2), we reconsider the domain of interest in terms of $\gamma$ units.

$$\Delta \gamma_{na} = \dot{\gamma} T_{na} \approx 255.6 \, \text{Hz} \times 1.819 \, \text{ms} = 0.465, \quad (1)$$
$$\Delta \gamma_{nl} = \dot{\gamma} T_{nl} \approx 255.6 \, \text{Hz} \times 1.113 \, \text{ms} = 0.284, \quad (2)$$

---
\* wingmc@fnal.gov

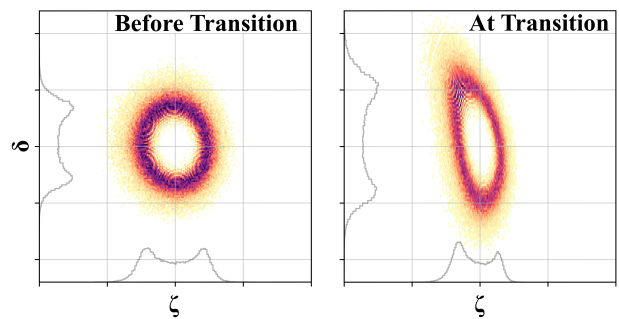

Figure 1: Longitudinal phase-space distributions in arbitrary units for the MI lattice using a predefined particle ensemble. The left distribution is captured 5.25 ms before transition, and the right distribution is captured at transition.

where $\dot{\gamma}$ reflects the ramp rate at the time of transition. This transformation simplifies the process of gauging $\gamma_t$ mitigation strategies since related beam instabilities can arise at any point during acceleration when $|\Delta \gamma| = |\gamma_t - \gamma| < \dot{\gamma} T_c$.

The assessed $\gamma_t$-jump scheme relies on the relationship between $\gamma_t$ and the lattice. By powering additional quadrupole magnets inserted into the MI, we can alter the value of $\gamma_t$ and manipulate $|\Delta \gamma|$ to reduce the durations of the instability envelopes and thus energy losses.

## XSUITE SIMULATION STUDIES

Using quadrupoles to control $\gamma_t$ is not a novel concept. The jump scheme selected for this analysis is based on Refs. [6, 7], and both sources predate the MI's construction. While this upgrade method promises improvements in beam efficiency, care must be taken to satisfy constraints with the jump-magnet placements so as not to detrimentally perturb the tunes, dispersions, or betas ($\beta_x$, $\beta_y$).

Four viable stations, each consisting of two focusing and defocusing quadrupoles, have been identified at locations around the MI ring. The defocusing magnets need to be installed in the straight sections where the dispersion magnitude is minimized, and focusing elements should be placed in the arcs with a $\pi$ phase advance established between the $\gamma_t$-jump elements. In the ideal case, the values of $\beta_x$ and $\beta_y$ should be the same across the magnets in a set.

Figure 2 displays the proposed magnet locations for the $\gamma_t$-jump system injected into the Xsuite lattice, assuming quadrupole lengths of 21.59 cm. The Courant-Snyder (CS) parameters for these positions are included in Table 1. The four stations are separated into two groups that reflect the MI's symmetry. Across two adjoined stations, the jump magnets are arranged in an F-F-DD-DD-F-F order, where F indicates a focusing magnet, and DD designates two adjacent defocusing quadrupoles. Given space constraints, the jump magnets could not be placed in the ideal positions, as evidenced by the slight $\beta$ mismatches, and therefore, the robustness of the modified lattice was tested through the implementation of specific errors in simulation sets.

Table 1: CS parameter table for the $\gamma_t$-jump magnets. $s_{End}$ is the downstream end of the component position along the beamline in meters. $\beta_{x,y}$ are the beta function values at $s_{End}$ in meters, and $\mu_{x,y}$ are the phase advances in units of $2\pi$. The line separators divide the magnets into station sets.

| Type | $s_{End}$ | $\beta_x$ | $\beta_y$ | $\mu_x$ | $\mu_y$ |
|---|---|---|---|---|---|
| F  | 452.26  | 52.09 | 12.07 | 3.71  | 3.57  |
| F  | 521.80  | 49.02 | 12.94 | 4.22  | 4.06  |
| DD | 573.77  | 49.82 | 12.97 | 4.72  | 4.56  |
| DD | 639.42  | 53.66 | 11.59 | 5.22  | 4.99  |
| F  | 691.10  | 52.36 | 11.68 | 5.73  | 5.48  |
| F  | 760.66  | 54.46 | 11.05 | 6.24  | 5.98  |
| F  | 2111.97 | 51.99 | 12.10 | 16.92 | 16.28 |
| F  | 2181.51 | 48.94 | 12.98 | 17.43 | 16.77 |
| DD | 2233.48 | 49.76 | 13.00 | 17.94 | 17.26 |
| DD | 2298.61 | 51.19 | 12.22 | 18.43 | 17.70 |
| F  | 2350.71 | 51.67 | 11.82 | 18.94 | 18.19 |
| F  | 2420.17 | 53.49 | 11.25 | 19.45 | 18.68 |

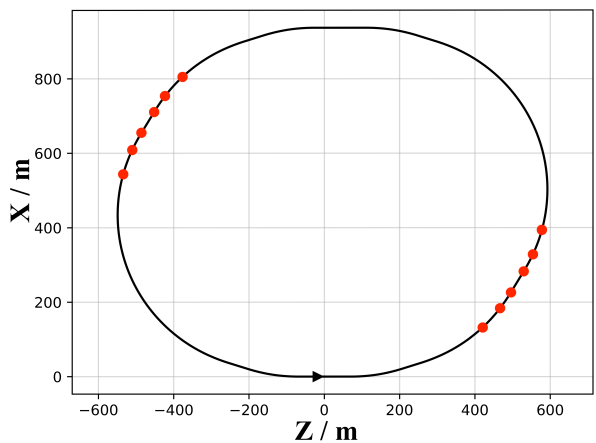

Figure 2: Xsuite line survey plot of the MI ring. The red circles indicate the locations of the $\gamma_t$-jump magnets.

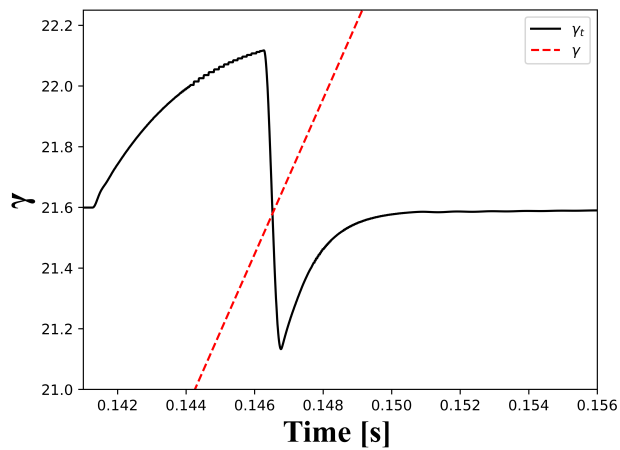

Figure 3: $\gamma_t(t)$ (*black, solid*) and particle $\gamma$ (*red, dashed*) around transition crossing in the Main Injector.

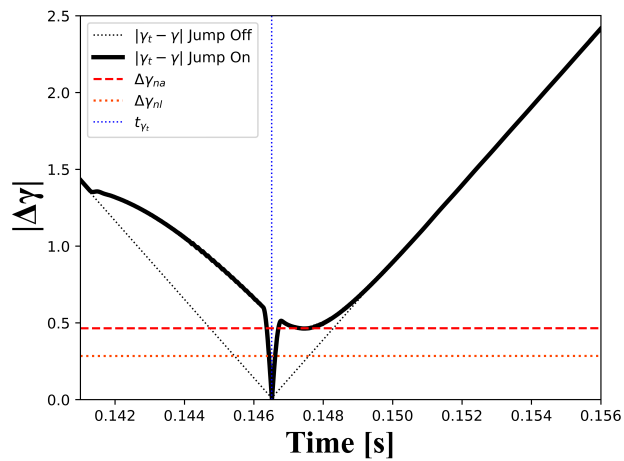

Figure 4: The black solid line shows $|\Delta\gamma(t)| = |\gamma_t - \gamma|$ with the jump quads powered. The dotted-black line displays the stock-lattice case. The horizontal lines indicate the nonadiabatic (*dashed*) and nonlinear (*dotted*) instability envelopes, and the vertical line depicts $t_{\gamma_t}$.

Returning to the fundamentals of the $\gamma_t$ jump, we know that the quads need to displace $\gamma_t$ by more than the value in Eq. (1) to avoid the instability envelopes. We also know there is an upper limit on the jump-magnet strength due to the $\beta$s, and we met these restrictions by applying bipolar pulses to the stations that yield peak integral fields of 0.85 T. The $\gamma_t$ jump is induced by setting the zero crossing of the pulse to $t_{\gamma_t}$ and by mapping the pulse to time-dependent assignments of the jump-magnets' quadrupole coefficients ($k_1(t)$).

Figure 3 demonstrates the changes induced on the lattice by powering the jump quadrupoles. $\gamma_t$ now shifts according to the bipolar waveform and meets a $\Delta\gamma_t = 0.5$ criterion before the required 0.5-ms polarity flip. In Fig. 4, we extract the viability of the jump scheme by examining $|\gamma_t - \gamma|$ in the time window around transition. The two horizontal lines depict the $\Delta\gamma$ envelopes associated with nonadiabatic (*dashed*) and nonlinear (*dotted*) instabilities. By applying the proposed jump scheme, the durations of the instability windows decrease by over 90%.

All simulations use an initial set of $10^5$ protons that emulate the injected distribution with a $20\pi$ mm mrad transverse emittance. Chromaticity ramps were also added to the momentum ramp to mimic MI operations. We first consider the cases of the stock MI lattice and the $\gamma_t$-jump scheme without additional uncertainties. From Fig. 5, we observe how the jump system improves the longitudinal phase space at $t_{\gamma_t}$ by reducing momentum smearing, spatial compression, and perturbation of the distribution's core.

With these qualitative checks completed, we focused on $E_{loss}$ as a quantitative evaluator of the scheme's robustness under various systematic scenarios. These results are summarized in Table 2, normalized to the PIP-II intensity, with $E_{loss}$ tabulated whenever a particle either struck the transverse aperture or crossed a $|\zeta| > 5.65$ m longitudinal cut at times $t - t_{\gamma_t} > -13.4$ ms. In the case of no errors, the simulation predicted a 95.71% reduction in $E_{loss}$ with the jump system in use.

Table 2: Summary of the expected $\gamma_t$-jump system performance under various scenarios. Applied uncertainties are described in the second and third columns, including magnet-to-magnet (M-M) and power-supply (I) effects. The $E_{loss}$ column lists the beam loss in terms of Joules/ramp, and the rightmost column compares the $E_{loss}$ value to the jump-off case.

| Mode | MI Quad Error | Applied $\gamma_t$-Jump Errors | $E_{loss}$ [J/ramp] | $E_{loss}$ Reduction |
|---|---|---|---|---|
| Stock, Jump Off | None | None | 57.20 | — |
| $\gamma_t$ Jump On | None | None | 2.45 | 95.71% |
| $\gamma_t$ Jump On | None | $\pm 0.2\%_{M-M} \pm 0.1\%_I \pm 24\,\mu s$ | 3.75±1.59 | 93.45%±2.79% |
| $\gamma_t$ Jump On | None | $\pm 0.5\%_{M-M} \pm 0.1\%_I \pm 24\,\mu s$ | 4.38±1.74 | 92.34%±3.04% |
| Preloaded Fluctuations | ±0.05% | $\pm 0.2\%_{M-M} \pm 0.1\%_I \pm 24\,\mu s$ | Variable | 65.60%±24.79% |

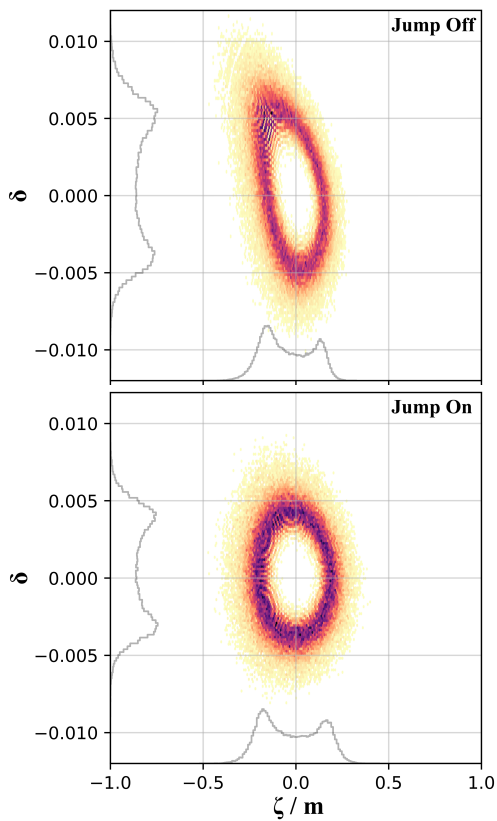

Figure 5: Longitudinal phase-space distributions at transition from Xsuite for the MI lattice. The top distribution depicts the stock case, and the bottom distribution shows the result with the jump system implemented.

Several studies probed magnet-to-magnet variations between the $\gamma_t$ quadrupoles, current fluctuations between the four magnet sets, and timing jitter between the supply triggers. For magnet-to-magnet variations, each of the $\gamma_t$ magnets had its $k_1(t)$ function multiplied by a random scalar between $(1 \pm \varepsilon_{M-M})$, where $\varepsilon_{M-M}$ is a preset value such as 0.2%. For current fluctuations and timing jitter, the $k_1(t)$ array was duplicated four times so independent arrays could be assigned to each magnet string. All values in these arrays received a random scaling factor between $(1 \pm \varepsilon_I)$, where $\varepsilon_I = 0.1\%$, and the arrays were then rolled by a random value between $\pm 24\,\mu s$. 100 simulations were produced for each studied case in Table 2.

Given the dependence on the $\beta$ functions, we also generated random $k_1$ scaling factors between $(1 \pm 0.05\%)$ for the MI quadrupoles. 100 different fluctuation sets were saved so the same variations could be evaluated with the jump magnets on and off.

We found no statistically significant difference between the no-error case and instances with $\varepsilon_{M-M} = 0.2\%$, $\varepsilon_I = 0.1\%$, and the timing jitter applied. The expected $E_{loss}$ shifts to 3.75±1.59 J/ramp. This corresponds to a significance of $0.81\sigma$ and an energy-loss reduction of 93.45%. We first noted signs of distinguishability at $\varepsilon_{M-M} = 0.5\%$. The MI-quad errors generate a large range of energy losses. However, the jump system always produced an $E_{loss}$ reduction at the level of 65.60%±24.79%.

## CONCLUSION

We evaluated the expected performance of a $\gamma_t$-jump system in the Fermilab MI using Xsuite. The analyzed lattice reduces the durations of both nonadiabatic and nonlinear instability windows by over 90%. In the absence of uncertainties, applying a bipolar pulse to 21.59-cm-long jump quadrupoles yielded a reduction in beam losses through transition of 95.71%.

With regard to the $\gamma_t$-jump magnets, we confirmed the viability of the ±0.2% magnet-to-magnet variability, ±0.1% current-fluctuation, and ±24 μs timing-jitter allowances. To assess MI ring quadrupole variations, 100 predefined random fluctuations altered the $k_1$ values of the MI quadrupole magnets by ±0.05%. Applying the same variations to simulations with the jump system on and off, we predicted that the jump-on case would reduce beam losses through transition by approximately 66%.

## ACKNOWLEDGMENTS

We thank Fermi National Accelerator Laboratory (Fermilab), a U.S. Department of Energy, Office of Science, HEP User Facility, for the resources provided. Fermilab is managed by the Fermi Forward Discovery Group LLC (FFDG), acting under Contract No. 89243024CSC000002. We also thank the FNAL EAF for prompt computation.